\documentclass[prx,twocolumn,floatfix,a4paper,superscriptaddress]{revtex4}

\usepackage{bm,color,amsmath,txfonts}
\usepackage{graphicx}
\usepackage{siunitx}
\usepackage{subfigure}
\usepackage{verbatim}
\usepackage{dcolumn}
\usepackage{bm}
\usepackage{epsf}
\usepackage{xcolor}
\usepackage{hyperref}
\usepackage{hhline}
\usepackage{float}
\usepackage{enumerate}
\usepackage{bbm}
\usepackage{lipsum}

\newcommand{\GG}{{\cal G}}

\begin{document}

\title{Stationary optomagnonic entanglement and magnon-to-optics quantum state transfer via opto-magnomechanics}

\author{Zhi-Yuan Fan}
\affiliation{Interdisciplinary Center of Quantum Information, State Key Laboratory of Modern Optical Instrumentation, and Zhejiang Province Key Laboratory of Quantum Technology and Device, School of Physics, Zhejiang University, Hangzhou 310027, China}
\author{Hang Qian}
\affiliation{Interdisciplinary Center of Quantum Information, State Key Laboratory of Modern Optical Instrumentation, and Zhejiang Province Key Laboratory of Quantum Technology and Device, School of Physics, Zhejiang University, Hangzhou 310027, China}
\author{Jie Li}\thanks{jieli007@zju.edu.cn}
\affiliation{Interdisciplinary Center of Quantum Information, State Key Laboratory of Modern Optical Instrumentation, and Zhejiang Province Key Laboratory of Quantum Technology and Device, School of Physics, Zhejiang University, Hangzhou 310027, China}

\begin{abstract}
We show how to prepare a steady-state entangled state between magnons and optical photons in an opto-magnomechanical configuration, where a mechanical vibration mode couples to a magnon mode in a ferrimagnet by the dispersive magnetostrictive interaction, and to an optical cavity by the radiation pressure. We find that, by appropriately driving the magnon mode and the cavity to simultaneously activate the magnomechanical Stokes and the optomechanical anti-Stokes scattering, a stationary optomagnonic entangled state can be created. We further show that, by activating the magnomechanical state-swap interaction and subsequently sending a weak red-detuned optical pulse to drive the cavity, the magnonic state can be read out in the cavity output field of the pulse via the mechanical transduction. The demonstrated entanglement and state-readout protocols in such a novel opto-magnomechanical configuration allow us to optically control, prepare, and read out quantum states of collective spin excitations in solids, and provide promising opportunities for the study of quantum magnonics, macroscopic quantum states, and magnonic quantum information processing.
\end{abstract}

\maketitle

\section{Introduction}
 
The past decade has witnessed a significant development and the formation of cavity magnonics~\cite{Hill,Naka19,YiLi,Yuan,RMP}. One of the pronounced advantages of the system is that the collective magnetic excitations (magnons) exhibit an excellent ability to interact coherently with a variety of distinct systems, including microwave photons~\cite{S1,S2,S3,S4,S5,S6}, optical photons~\cite{Usami16,Tang16,Haigh,PRB16,Usami18,Zhu,Haigh21}, vibration phonons~\cite{SA,Jie18,Jie20,Davis,Jie22}, and superconducting qubits~\cite{qubit1,qubit2,qubit3}, etc. The composite architecture based on magnons promises broad application prospects in quantum information processing, quantum sensing, and quantum networks~\cite{Naka19,YiLi,Yuan,JiePRX}. In particular, the coupling between magnons and microwave or optical photons offers the possibility to optically control, engineer, detect, and transmit magnonic states~\cite{S1,S2,S3,S4,S5,S6,Usami16,Tang16,Haigh,PRB16,Usami18,Zhu,Haigh21,OM3,OM4,OM5,OM6,OM7,OM8,OM8b,OM9,OM10,OM11,OM12,OM13,OM14}. This coupling is an indispensable component to build a magnonic quantum network with remote quantum nodes connected by light~\cite{JiePRX}. The mature and handy optical means also allow us to prepare non-classical magnonic states~\cite{OM3,OM4,OM5,OM8b,OM9,OM10,OM11}.

Among a variety of non-classical states, entangled states find a particularly wide range of applications, e.g., in quantum information processing~\cite{en2,en3}, quantum teleportation~\cite{tele}, quantum metrology~\cite{metro}, and fundamental tests of quantum mechanics~\cite{bell,Grav}. Magnons and microwave cavity photons can get entangled by exploiting the nonlinear magnetostrictive coupling to vibration phonons~\cite{Jie18,Jie19}. Alternatively, they can also be entangled by utilizing the dissipative coupling between magnons and photons~\cite{Yuan20}. In the optical frequency, magnons and optical photons can be entangled by using optical pulses to active the optomagnonic Stokes scattering~\cite{OM4,OM9,OM10,OM11,JiePRX}. Typical optomagnonic systems, either a whispering-gallery-mode~\cite{Usami16,Tang16,Haigh}, or a waveguide~\cite{Zhu}, or a microcavity~\cite{Haigh21} configuration, suffer a significantly large cavity decay rate $\kappa_a$ (typically in gigahertz) and a much smaller effective optomagnonic coupling strength $G_{\rm om} \ll \kappa_a$, thus putting the system well within the weak-coupling regime. Therefore, in all of the aforementioned protocols~\cite{OM4,OM9,OM10,OM11,JiePRX}, fast optical pulses are adopted to generate a {\it transient} optomagnonic entangled state. This evades the stringent requirement on the optomagnonic cooperativity ${\cal C}_{\rm om} = \frac{G_{\rm om}^2}{\kappa_a \kappa_m} > 1$ ($\kappa_m$ is the magnon dissipation rate) required for preparing a {\it stationary} entangled state. 

Here we provide a novel approach to preparing a stationary entangled state between magnons and optical photons by using a mechanical vibration mode being an intermediary, which couples to the magnons and the photons by the nonlinear magnetostrictive and radiation-pressure interaction, respectively. This indirect-coupling configuration~\cite{Fan} takes advantage of the low mechanical damping rate $\gamma_b$, and can easily achieve both the magnomechanical and optomechanical cooperativities ${\cal C}_{\rm mM} = \frac{G_{\rm mM}^2}{\kappa_m \gamma_b} > 1$, and ${\cal C}_{\rm oM} = \frac{G_{\rm oM}^2}{\kappa_a \gamma_b} > 1$, under current technology~\cite{SA,Davis,Jie22,MA}. This enables us to generate a {\it stationary} optomagnonic entangled state without requiring the stringent condition ${\cal C}_{\rm om} >1$ in current directly-coupled optomagnonic systems. Specifically, the magnons and optical photons are prepared in a steady entangled state by simultaneously activating the magnomechanical Stokes and the optomechanical anti-Stokes scattering, realized by properly driving the magnon mode and the optical cavity. The steady-state optomagnonic entanglement is more useful as it can provide a stable entanglement source to be applied to the quantum information science.
	
We further show that, by activating the magnomechanical state-swap operation realized by driving the magnon with a red-detuned microwave field, the magnonic state can be transferred to the mechanical mode. We then switch off the microwave drive and, after a short period, send a weak red-detuned optical pulse to the cavity to activate the optomechanical beam-splitter interaction. The magnonic state is then read out in the output field of the pulse via the mechanical transduction. To give a concrete example, we show that the magnon squeezing can be transferred to the optical pulse with high fidelity. This offers an efficient optical means to read out localized magnonic states in solids.

The paper is organized as follows. In Sec.~\ref{OMM}, we introduce the opto-magnomechanical system, which is used for our entanglement and state-readout protocols. We strictly derive the Hamiltonian of the tripartite system and, in particular, the magnomechanical dispersive interaction Hamiltonian, which provides essential nonlinearity for producing the entanglement. We then show how to use this hybrid system to obtain stationary optomagnonic entanglement in Sec.~\ref{entangle}, and how to use optical pulses to read out the magnonic states in Sec.~\ref{read}. Finally, we summarize the results in Sec.~\ref{conc}.

\section{Opto-magnomechanical system}\label{OMM}

The magnon-phonon dispersive coupling is a vital component of the entanglement protocol. It provides essential nonlinearity for creating the magnomechanical entanglement~\cite{Jie18}, which gives rise to the optomagnonic entanglement if the state-swap operation between the phonons and the photons is activated. In Sec.~\ref{MMc}, we show how to quantize the magnetization, the strain displacement, and the magnetoelastic energy. This allows us to strictly derive the Hamiltonian of the magnomechanical dispersive interaction, and specify the condition under which such a dispersive coupling is dominant in the magnetoelastic coupling. We then introduce, in Sec.~\ref{model}, the complete model, namely the opto-magnomechanical system, that is adopted in the entanglement and state-readout protocols.

\subsection{Magnomechanical coupling}\label{MMc}

The magnetoelastic coupling describes the interaction between the magnetization and the elastic strain of the magnetic material. There are three kinds of interactions, depending on the distance between magnetic atoms (or ions): The spin-orbital interaction, the exchange interaction, and the magnetic dipole-dipole interaction~\cite{Gurevich}.  For a cubic crystal, the magnetoelastic energy density is given by~\cite{Kittel49}
		\begin{align}\label{fme}
		\begin{split}
			f_{\rm me}=&\frac{B_1}{M_S^2} \left( M_x^2\epsilon_{xx}+M_y^2\epsilon_{yy}+M_z^2\epsilon_{zz} \right)\\
			&+\frac{2B_2}{M_S^2} \left( M_xM_y\epsilon_{xy}+M_xM_z\epsilon_{xz}+M_yM_z\epsilon_{yz} \right),
		\end{split}
		\end{align}
where $B_1$ and $B_2$ are the magnetoelastic coupling coefficients, $M_S$ is the saturation magnetization, and $M_{x,y,z}$ are the magnetization components. $\epsilon_{ij}=\frac{1}{2}\left ( \partial u_i/\partial l_j +\partial u_j/\partial l_i\right )$ denotes the strain tensor of the magnetic crystal, with coordinate indices $i,j\in \left\{ x,y,z\right\}$, and $u_{x,y,z}$ are the components of the displacement vector $\vec{u}$.
	
		
Using the Holstein-Primakoff transformation~\cite{HPT}, the magnetization can be quantized as 
 \begin{equation}
  m= \sqrt{\frac{V}{2\hbar\gamma M_S}}\left (M_x - i M_y \right),
 \end{equation}
 where $m$ denotes the magnon mode operator, $V$ is the volume of the crystal, and $\gamma$ is the gyromagnetic ratio. Thus,  we obtain
		\begin{align}
		\begin{split}
			M_x=&\sqrt{\frac{\hbar\gamma M_S}{2V}}\left( m+m^\dagger \right), \\
			M_y=&i\sqrt{\frac{\hbar\gamma M_S}{2V}} \left( m-m^\dagger \right), \\
		\end{split}
		\end{align}
		and 
		\begin{align}
		M_z=\left( M_S^2-M_x^2-M_y^2 \right)^{\frac{1}{2}}\simeq  M_s-\frac{\hbar \gamma}{V}m^\dagger m.  
		\end{align} 
Substituting the above expressions into the magnetoelastic energy density $f_{\rm me}$ and integrating over the whole volume of the crystal, we obtain the semiclassical magnetoelastic Hamiltonian. The first line in Eq.~\eqref{fme} gives rise to
		\begin{align}\label{H1}
		\begin{split}
			H_1\simeq\ &\frac{B_1}{M_S}\frac{\hbar\gamma}{V}m^\dagger m\int dl^3 \left( \epsilon_{xx}+\epsilon_{yy}-2\epsilon_{zz} \right)\\
			&+ \frac{B_1}{M_S}\frac{\hbar\gamma}{2V} \left(m^2+m^{\dagger 2} \right) \int dl^3 \left(\epsilon_{xx}-\epsilon_{yy} \right)\\
			&+\frac{B_1}{M_S^2}\frac{\hbar^2\gamma^2}{V^2}m^\dagger m m^\dagger m \int dl^3\epsilon_{zz}\ ,
		\end{split}
		\end{align}
		and the second line yields the Hamiltonian
		\begin{align}\label{H2}
		\begin{split}
			H_2\simeq & \ 
			i\frac{B_2}{M_S}\frac{\hbar\gamma}{V} \left( m^2-m^{\dagger 2} \right)\int dl^3 \epsilon_{xy}
			+\frac{2B_2}{M_S^2}\sqrt{\frac{\hbar\gamma M_S}{2V}}\\
			&\times\left ( M_s-\frac{\hbar\gamma}{V}m^\dagger m \right )\left [ m\int dl^3 \left(\epsilon_{xz}+i\epsilon_{yz} \right)+\textup{H.c.} \right ].
		\end{split}
		\end{align}
The magnetoelastic Hamiltonian, $H_{\rm me}=H_1+H_2$, implies diverse magnon-phonon interactions, depending on their frequency relation. For given magnon and phonon frequencies, a certain type of coupling can be dominant in the magnetoelastic coupling, while other coupling mechanisms play only a negligible role. This can be seen more clearly in the fully quantized interaction Hamiltonian by further quantizing the strain displacement.


The magnetoelastic displacement can be decomposed and expressed as the superposition form of
		\begin{align}
		\vec{u}(x,y,z)=\sum_{n,m,k}d^{(n,m,k)}\vec{\chi}^{(n,m,k)}(x,y,z),
		\end{align}
where $\vec{\chi}^{(n,m,k)}(x,y,z)$ represents the dimensionless displacement eigenmode, and $d^{(n,m,k)}$ is the corresponding amplitude, with the mode indices $(n,m,k)$. The displacement amplitude $d^{(n,m,k)}$ can be quantized as
		\begin{align}
		d^{(n,m,k)}=d_{\textup{zpm}}^{(n,m,k)}\left(b_{n,m,k}+b^\dagger_{n,m,k} \right),
		\end{align}
where $d_{\textup{zpm}}^{(n,m,k)}$ denotes the amplitude of the zero-point motion, and $b_{n,m,k}$ ($b^\dagger_{n,m,k}$) is the annihilation (creation) operator of the corresponding phonon mode.

By quantizing the strain displacement in the semiclassical Hamiltonians of Eqs.~\eqref{H1} and \eqref{H2}, the fully quantized magnetoelastic Hamiltonian can be derived, which accounts for diverse magnon-phonon interactions when their frequencies vary. To be specific, when the phonon frequencies are much lower than the magnon frequency, $\omega_b^{(n,m,k)}\ll \omega_m$ (typically the case for a large-size crystal~\cite{SA,Davis,Jie22}), and by neglecting the fast-oscillating terms, we obtain the dominant dispersive-type interaction
		\begin{align}\label{disp}
		H_{\rm me} \simeq \sum_{n,m,k}\hbar g_{\textup{disp}}^{(n,m,k)}m^\dagger m \, \left(b_{n,m,k}+b^\dagger_{n,m,k} \right),
		\end{align}
where the dispersive coupling strength 
		\begin{align}
		\begin{split}
			g_{\textup{disp}}^{(n,m,k)}=&\frac{B_1}{M_S}\frac{\gamma}{V}\int dl^3\ d_{\textup{zpm}}^{(n,m,k)}\\
			&\times \left ( \frac{\partial \chi_x^{(n,m,k)}}{\partial x} +\frac{\partial \chi_y^{(n,m,k)}}{\partial y}-2\frac{\partial \chi_z^{(n,m,k)}}{\partial z} \right ). 
		\end{split}
		\end{align}
Note that the above Hamiltonian is derived under the condition of low-lying magnon excitations, where the second-order term of the magnon excitation $m^\dag m$ (i.e., the third line in Eq.~\eqref{H1}) can be safely neglected.

For the case where the magnon and phonon frequencies are close, $\omega_b^{(n,m,k)} \, {\simeq}\,\, \omega_m$, by neglecting the nonresonant fast-oscillating terms, we get the following linear interaction Hamiltonian:
	\begin{align}
		H_{\rm me} \simeq \sum_{n,m,k}\hbar \left(  g_{\textup{lin}}^{(n,m,k)}m b^\dagger_{n,m,k} +\textup{H.c.} \right) ,
	\end{align}
		where the linear coupling rate
	\begin{align}
		\begin{split}
			g_{\textup{lin}}^{(n,m,k)}=&\frac{B_2}{M_S}\sqrt{\frac{\gamma M_S}{2\hbar V}}\ \times \\
			&\left [ \int dl^3\ d_{\textup{zpm}}^{(n,m,k)}
			\left ( \frac{\partial \chi_x^{(n,m,k)}}{\partial z}+\frac{\partial \chi_z^{(n,m,k)}}{\partial x} \right )\right. \\
			&\left. +\ i\int dl^3\ d_{\textup{zpm}}^{(n,m,k)}\left ( \frac{\partial \chi_y^{(n,m,k)}}{\partial z}+\frac{\partial \chi_z^{(n,m,k)}}{\partial y} \right )\right ] .
		\end{split}
	\end{align} 
Similarly, we have neglected the higher-order term under the low-lying excitations. This magnon-phonon linear coupling is widely used for gigahertz phonons, and the strong coupling can be achieved forming the magnon-phonon polaron~\cite{strong}. However, such a beam-splitter type coupling cannot create entanglement, and therefore is not desired for our protocol.

Apart from the dispersive and linear couplings discussed above, {\it both} the magnetoelastic Hamiltonian $H_1$ and $H_2$ imply the magnon parametric amplification (PA) if the phonon frequency is twice of the magnon frequency, $\omega_b^{(n,m,k)} \, {\simeq}\,\, 2\omega_m$. In this situation, we obtain the following interaction Hamiltonian by neglecting nonresonant fast-oscillation terms:
	\begin{align}
		H_{\rm me} \simeq\sum_{n,m,k}\hbar \left( g_{\textup{PA}}^{(n,m,k)}m^2 b_{n,m,k}^\dagger+\textup{H.c.} \right),
	\end{align}
		where the corresponding coupling rate is given by
	\begin{align}
		\begin{split}
			g_{\textup{PA}}^{(n,m,k)}&=\frac{1}{M_S}\frac{\gamma}{2V} \int dl^3 \ d_{\textup{zpm}}^{(n,m,k)} \times \\
			& \left [ B_1\left( \frac{\partial \chi_x^{(n,m,k)}}{\partial x}-\frac{\partial \chi_y^{(n,m,k)}}{\partial y} \right ) + i\ B_2 \left ( \frac{\partial \chi_x^{(n,m,k)}}{\partial y}+\frac{\partial \chi_y^{(n,m,k)}}{\partial x} \right )\right ].
		\end{split}
	\end{align}
Such magnon PA can be used to generate entangled pairs of magnons, in analogy to the generation of entangled photon pairs by optical parametric down-conversion.

To sum up, in this section we provide a strict derivation of the dominant effective magnomechanical Hamiltonian for different situations of the magnon and phonon frequencies: Specifically, {\it i}) a dispersive coupling for lower-frequency phonons, $\omega_b^{(n,m,k)}\ll \omega_m$; {\it ii}) a linear coupling for nearly resonant magnons and phonons, $\omega_b^{(n,m,k)}\simeq \omega_m$; and {\it iii}) a magnon PA coupling for the phonon frequency being twice of the magnon frequency, $\omega_b^{(n,m,k)}\simeq 2\omega_m$. 
The effective Hamiltonian is obtained by neglecting inappreciable fast-oscillating terms, which is a good approximation when the magnon frequency and (or) the phonon frequency are (is) much larger than their coupling strength and dissipation rates.
		

 \begin{figure}[b]
	\includegraphics[width=0.8\linewidth]{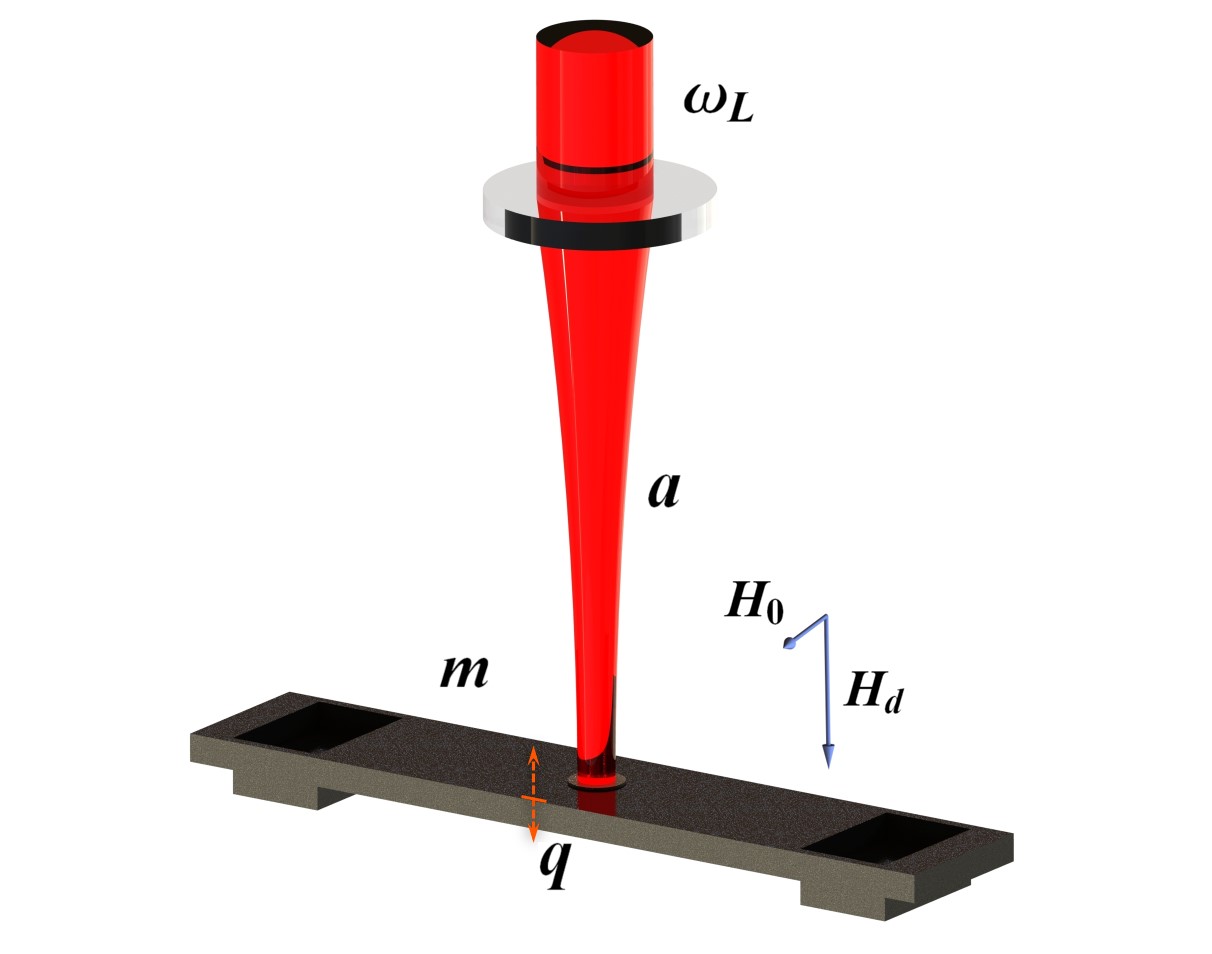}
	\caption{Sketch of the opto-magnomechanical system. The mechanical displacement couples to the magnon excitations in a YIG micro bridge via the dispersive magnetostriction interaction, and to an optical cavity via the radiation-pressure interaction.}
	 \label{fig1}
	 \end{figure}

\subsection{Hamiltonian and Langevin equations of opto-magnomechanics}\label{model}

The derivation of the magnomechanical Hamiltonian indicates that the mechanical frequency should be much lower than the magnon frequency in order to have a dispersive interaction. The magnomechanical system can be a yttrium-iron-garnet (YIG) three-dimensional magnon nanoresonator~\cite{PRAp}, in which long-lived spin wave excitations are coupled to mechanical vibrations. In Ref.~\cite{PRAp}, a micron-sized YIG bridge supports a magnon mode with the frequency in gigahertz and a mechanical vibration mode with the frequency ranging from tens to hundreds of megahertz. Therefore, such a system possesses a dominant magnon-phonon dispersive coupling. Another advantage of the system lies in the narrow linewidth (a few megahertz) of the magnetic excitations. 

By attaching a small high-reflectivity mirror pad to the surface of the YIG bridge, as depicted in Fig.~\ref{fig1}, the magnetostriction induced mechanical displacement can be coupled to an optical field via radiation pressure~\cite{Dirk,SG,EP}, forming an optomechanical cavity with another fixed mirror. Such a tripartite opto-magnomechanical configuration was adopted to optically measure the magnon population~\cite{Fan}. We note that the attached mirror pad should be fabricated {\it sufficiently small and light}, such that it will not appreciably affect the mechanical properties (e.g., the mechanical displacement and damping) of the micro bridge. The deformation displacement can be approximately regarded as uniform in the direction perpendicular to the attached surface (with negligible bending displacement), such that the YIG bridge and the mirror can stick together tightly and are integrated into one body, which oscillate approximately with the same frequency.  Alternatively, one may consider using the ``membrane-in-the-middle" configuration~\cite{Jack} by placing the YIG sample in the middle of an optical cavity. In this configuration, the magnomechanically induced displacement can also couple dispersively to the optical cavity.

The Hamiltonian of the hybrid opto-magnomechanical  system reads
\begin{align}
	\begin{split}
		H/\hbar =& \! \sum_{j=m,a} \omega_j j^\dagger j +\frac{\omega_b}{2}\left ( q^2+p^2 \right ) + g_m m^\dagger m q - g_a a^\dagger a q \\
		 +& i\Omega\left ( m^\dagger e^{-i\omega_0 t}-m e^{i\omega_0 t} \right ) + iE\left ( a^\dagger e^{-i\omega_L t}-a e^{i\omega_L t} \right ),
	\end{split}
\end{align}
where $a$ and $a^\dagger$ ($m$ and $m^\dagger$) are the annihilation and creation operators of the cavity (magnon) mode, satisfying the canonical commutation relation $[j,j^\dagger]=1$ ($j=a,m$), $q$ and $p$ denote the dimensionless mechanical position and momentum, $[q,p]=i$, and $\omega_k$ ($k=a,m,b$) are the resonance frequencies of the cavity, magnon, and mechanical modes, respectively. The magnon mode is activated by placing the YIG micro bridge in a uniform bias magnetic field ($H_0$) and applying a microwave field with its magnetic component ($H_d$) perpendicular to the bias field, see Fig.~\ref{fig1}. The magnon frequency is tunable by varying the bias magnetic field. $g_m$ ($g_a$) is the bare magnomechanical (optomechanical) coupling rate, and the effective coupling can be improved by driving the magnon (cavity) mode with a microwave (laser) field. The Rabi frequency associated with the microwave drive is $\Omega=\frac{\sqrt{5}}{4}\gamma \sqrt{N}H_d$ ($\gamma$: the gyromagnetic ratio; $N$: the total number of spins)~\cite{Jie18}, and the cavity-laser coupling strength is $E=\sqrt{\kappa_aP_L/(\hbar\omega_L)}$, with $P_L$ ($\omega_L$) being the power (frequency) of the laser and $\kappa_a$ being the cavity decay rate.

Including the dissipation and input noise of each mode and working in the interaction picture with respect to $\hbar \omega_L a^\dagger a+\hbar \omega_0 m^\dagger m$, we obtain the following quantum Langevin equations (QLEs):
\begin{align}\label{QLEs1}
	\begin{split}
	\dot{q}=&\ \omega_b p, \,\,\,\,  \dot{p}=-\omega_b q -\gamma_b p - g_m m^\dagger m + g_a a^\dagger a + \xi,\\
	  \dot{a}=&-i\Delta_a a-\frac{\kappa_a}{2}a+ig_a a q+E+\sqrt{\kappa_a}a_{in},  \\
        \dot{m}=&-i\Delta_m m-\frac{\kappa_m}{2}m-ig_m m q+\Omega+\sqrt{\kappa_m}m_{in},       
	\end{split}
\end{align}
where $\Delta_a = \omega_a - \omega_L$, $\Delta_m = \omega_m - \omega_0$, and $\gamma_b$, $\kappa_a$, and $\kappa_m$ are the dissipation rates of the mechanical, cavity, and magnon modes, respectively. The corresponding zero-mean input noise operators $\xi$, $a_{in}$, and $m_{in}$ obey the following correlation functions: $\left<\xi(t)\xi(t')+\xi(t')\xi(t) \right>/2\simeq \gamma_b \left[ 2N_b(\omega_b)+1 \right] \delta(t-t')$ under the Markovian approximation, which is valid for a large mechanical quality factor $Q=\omega_b/\gamma_b\gg 1$~\cite{DV}, $\langle a_{in}(t)a^\dagger_{in}(t') \rangle =\delta(t-t')$, $\langle m_{in}(t) \, m^\dagger_{in}(t') \rangle=\left[N_m(\omega_m)+1 \right] \delta(t-t')$, and $\langle m^\dagger_{in}(t) \, m_{in}(t') \rangle=N_m(\omega_m)\delta(t-t')$, where the mean thermal excitation number $N_j(\omega_j)=\left[ \textup{exp}\left ( \hbar\omega_j/k_B T \right )-1\right]^{-1}$ ($j=b,m$), with $k_B$ being the Boltzmann constant and $T$ being the temperature of the reservoir of the system.

Since the magnon and cavity modes are strongly driven, resulting in large steady-state amplitudes $\left|\left< m \right> \right|,\left|\left< a\right> \right|\gg 1$, the nonlinear opto- and magnomechanical dynamics can be linearized around large classical averages. This is realized by writing each mode operator as the sum of its classical average and a quantum fluctuation operator, $k=\left\langle k\right\rangle +\delta k\ (k=a,m,q,p)$, and neglecting small second-order fluctuation terms. Consequently, the QLEs \eqref{QLEs1} are separated into two sets of equations for classical averages and quantum fluctuations, respectively. By solving the former set of equations for the classical averages in the steady state, we obtain the following solutions:
	\begin{align}
		\begin{split}
		\left< q\right>&=\left( g_a\left|\left<a \right> \right|^2-g_m\left|\left< m\right> \right|^2 \right) /\omega_b,   \,\,\,\,\,\,  \left< p\right>=0, \\
		\left< a\right>&=\frac{E}{i\tilde{\Delta}_a+\frac{\kappa_a}{2}},\,\,\,\,\,\,\,   \left< m\right>=\frac{\Omega}{i\tilde{\Delta}_m+\frac{\kappa_m}{2}}, 
		\end{split}
	\end{align}
where $\tilde{\Delta}_a=\Delta_a - g_a \left<q \right>$  ($\tilde{\Delta}_m=\Delta_m+g_m\left<q \right>$) is the effective cavity (magnon)-drive detuning by including the frequency shift due to the mechanical displacement jointly caused by the opto- and magnomechanical interactions.
	
The linearized QLEs for the quantum fluctuations are given in the quadrature form by 
	 \begin{align}\label{QLEs2}
	 	\begin{split}
	 		\delta \dot{q}=&\ \omega_b \delta p, \\
	 		\delta \dot{p}=&-\omega_b \delta q-\gamma_b \delta p+\textup{Im}G_a \delta X_a-\textup{Re}G_a\delta Y_a\\
	 		& +\textup{Im}G_m\delta X_m-\textup{Re}G_m\delta Y_m+\xi,\\
	 		\delta \dot{X}_a=&-\frac{\kappa_a}{2}\delta X_a+\tilde{\Delta}_a\delta Y_a+\textup{Re}G_a\delta q+\! \sqrt{\kappa_a}X_{a,in}, \\
	 		\delta \dot{Y}_a=&-\tilde{\Delta}_a\delta X_a-\frac{\kappa_a}{2}\delta Y_a+\textup{Im}G_a\delta q+\! \sqrt{\kappa_a}Y_{a,in},\\
	 		\delta \dot{X}_m=&-\frac{\kappa_m}{2}\delta X_m+\tilde{\Delta}_m\delta Y_m+\textup{Re}G_m\delta q+\! \sqrt{\kappa_m}X_{m,in},\\
	 		\delta \dot{Y}_m=&-\tilde{\Delta}_m\delta X_m-\frac{\kappa_m}{2}\delta Y_m+\textup{Im}G_m\delta q+ \! \sqrt{\kappa_m}Y_{m,in},
	 	\end{split}
	 \end{align}
 where the quadrature fluctuations $\delta X_k=(\delta k+\delta k^\dagger)/\sqrt{2}$,\ $\delta Y_k=i(\delta k^\dagger - \delta k)/\sqrt{2}$, and the quadratures of the input noises $X_{k,in}=(k_{in}+k^\dagger_{in})/\sqrt{2}$, and $Y_{k,in}=i(k^\dagger_{in}-k_{in})/\sqrt{2}$ ($k=a,m$). The effective opto- and magnomechanical coupling strengths are $G_a=i\sqrt{2} g_a\left\langle a \right\rangle $, and $G_m=-i\sqrt{2}g_m\left\langle m \right\rangle$, respectively.
 
The QLEs \eqref{QLEs2} can be rewritten in a compact matrix form of
	\begin{align}
		\dot{u}(t)=Au(t)+n(t),
	\end{align}
where $u(t)$ is the vector of the quadrature fluctuations, $u(t) \,\,{=}\,\, [ \delta q(t), \, \delta p(t), \, \delta X_a(t), \, \delta Y_a(t), \, \delta X_m(t),\, \delta Y_m(t) ]^T$, $n(t)$ is the vector of the input noises, $n(t)=[0,\xi(t), \sqrt{\kappa_a}\delta X_{a,in}(t),$  $\sqrt{\kappa_a}\delta Y_{a,in}(t),\sqrt{\kappa_m}\delta X_{m,in}(t), \sqrt{\kappa_m}\delta Y_{m,in}(t) ] ^T$, and the drift matrix $A$ is given by	
\begin{equation}
A=\begin{pmatrix}
			0 & \omega_b & 0 & 0 & 0 & 0 \\
			-\omega_b & -\gamma_b & 0 & -G_a & 0 & -G_m \\
			G_a & 0 & -\frac{\kappa_a}{2} & \tilde{\Delta}_a  & 0 & 0 \\
			0 & 0 & -\tilde{\Delta}_a & -\frac{\kappa_a}{2} & 0 & 0 \\
			G_m & 0 & 0 & 0 & -\frac{\kappa_m}{2} & \tilde{\Delta}_m \\
			0 & 0 & 0 & 0 & -\tilde{\Delta}_m & -\frac{\kappa_m}{2} \\
\end{pmatrix}.
\end{equation}
Note that the above drift matrix is given under the optimal condition for the entanglement, $\left| \tilde{\Delta}_a \right|, \, \left| \tilde{\Delta}_m\right|  \simeq \omega_b \gg \kappa_a, \kappa_m$~\cite{Jie18,Jie20}. This gives rise to approximately pure imaginary steady-state amplitudes $\left<a \right>$ and $\left<m \right>$, and thus real opto- and magnomechanical coupling strengths $G_a$ and $G_m$. 
	
Under the condition $\omega_b \gg G, \kappa$, which validates the rotating wave approximation, the {\it dispersive} opto- and magnomechanical couplings enable the realization of distinct opto- and magnomechanical operations by properly driving the system, i.e., the parametric down-conversion for a blue-detuned drive, $\tilde{\Delta } \simeq -\omega_b$, and the beam-splitter (state-swap) operation for a red-detuned drive, $\tilde{\Delta } \simeq\omega_b$. These operations are the building blocks for realizing the two protocols to be discussed in the following sections.

\section{Stationary optomagnonic entanglement}\label{entangle}

\begin{figure}[b]
\includegraphics[width=\linewidth]{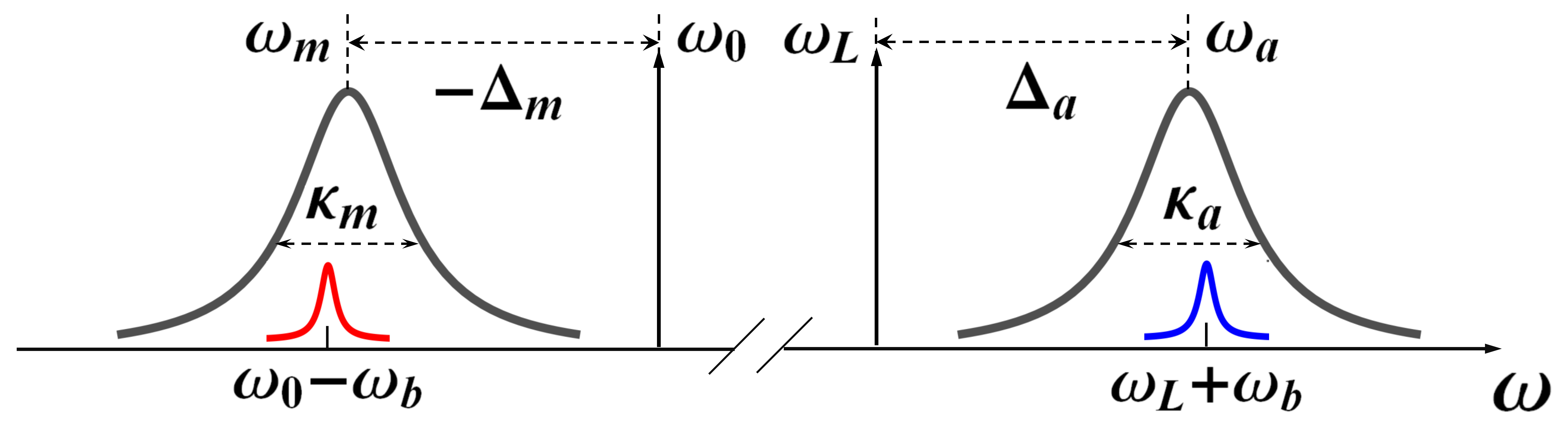}
\caption{Mode frequencies and linewidths adopted in the entanglement protocol.  The mechanical motion with the frequency $\omega_b > \kappa_{m,a}$ scatters both microwave and laser driving photons onto the Stokes and anti-Stokes sidebands at  $\omega_0 \mp \omega_b$, and $\omega_L \mp \omega_b$, respectively.  When the magnon (cavity) mode is resonant with the Stokes (anti-Stokes) sideband of the microwave (laser) drive field at the frequency $\omega_0$ ($\omega_L$), an entangled state between magnons and optical photons can be generated in the steady state. }
 \label{fig2}
\end{figure}

In this section, we show how to achieve the stationary entanglement between magnons and optical photons via the mediation of phonons. Specifically, we drive the magnon (cavity) mode with a blue-detuned microwave field (a red-detuned laser field), and the detuning is equal to the mechanical frequency, i.e., $-\tilde{\Delta}_m=\tilde{\Delta }_a=\omega_b$, such that the magnon (cavity) mode is resonant with the magnomechanical (optomechanical) Stokes (anti-Stokes) sideband, see Fig.~\ref{fig2}. We consider the case of resolved sidebands ($\omega_b > \kappa_{m,a}$), which can be achieved for a micron-sized YIG bridge~\cite{PRAp} and a typical optomechanical cavity~\cite{MA}.  The simultaneous activation of the magnomechanical parametric down-conversion (leading to magnon-phonon entanglement) and the optomechanical beam-splitter interaction (yielding photon-phonon state swapping) results in magnon-photon entanglement. The optomagnonic entanglement can be prepared in the steady state, because of the achievable large cooperativities ${\cal C}_{\rm mM}, {\cal C}_{\rm oM}  \gg 1$. 

We now show explicitly how the stationary optomagnonic entanglement can be achieved. For our system, due to the linearized dynamics and the Gaussian nature of the input noises, the steady state of the system quantum fluctuations is a three-mode Gaussian state, which can be fully characterized by the covariance matrix (CM) $V$, with the entries defined as $V_{ij}=\langle u_i(t)u_j(t')+u_j(t')u_i(t) \rangle/2$ $(i,j=1,2,...,6)$. The steady-state CM of the system can be obtained by directly solving the Lyapunov equation~\cite{DV07}
\begin{equation}
	AV+VA^T=-D,
\end{equation}
where $D\,\,{=}\,\,\textup{diag}\,\Big[ 0, 2\gamma_b(N_b{+}\frac{1}{2}), \frac{\kappa_c}{2},  \frac{\kappa_c}{2}, \kappa_m(N_m {+} \frac{1}{2}),\kappa_m(N_m {+} \frac{1}{2}) \Big]$ is the diffusion matrix, which is defined by $D_{ij}=\langle n_i(t)n_j(t')+n_j(t')n_i(t) \rangle/2$. The Gaussian bipartite entanglement is quantified by the logarithmic negativity~\cite{LN}, defined as
\begin{equation}
	E_N=\textup{max}\left[ 0,-\textup{ln}(2\eta^{-})\right],
\end{equation}
where $\eta^{-} \equiv 2^{-1/2}\, \big[ \Sigma - \big(\Sigma^2 - 4\, \textup{det}\, V_{4} \big)^{1/2}\big]^{1/2}$, and $V_{4}=\big[V_a,V_{am}; V_{am}^T,V_m \big]$ is the $4\times4$ CM associated with the optomagnonic subsystem, with $V_a,V_m,V_{am}$, and $V_{am}^T$ being the $2\times2$ blocks of $V_{4}$, and $\Sigma \equiv \textup{det}\,V_a+\textup{det}\,V_m-2\textup{det}\,V_{am}$.

\begin{figure}[t]
\includegraphics[width=\linewidth]{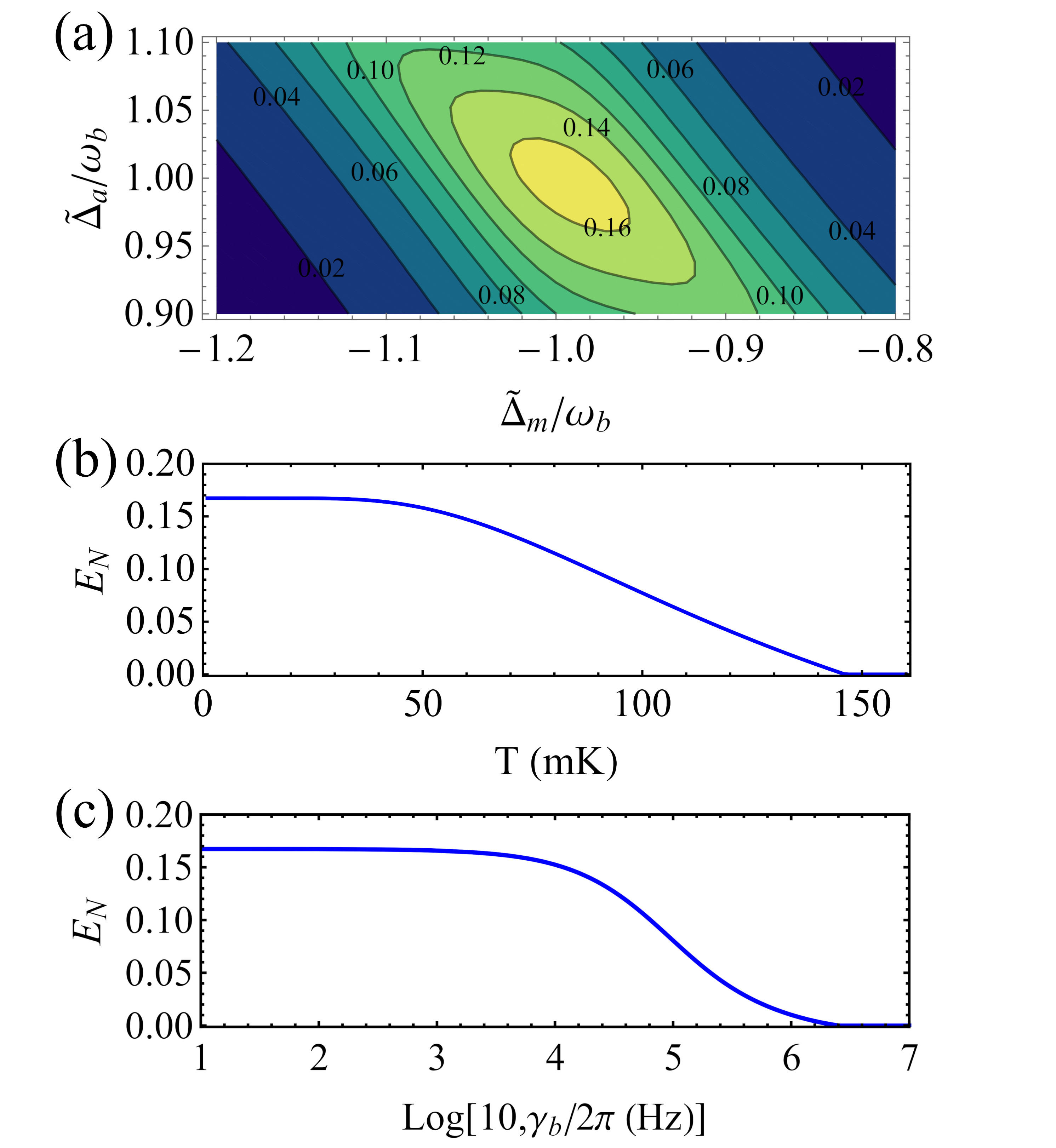}
\caption{(a) Contour plot of the stationary optomagnonic entanglement $E_N$ versus detunings $\tilde{\Delta}_m$ and $\tilde{\Delta}_a$. The entanglement $E_N$ versus (b) temperature $T$ and (c) mechanical damping rate $\gamma_b$ at the optimal detunings $-\tilde{\Delta }_m=\tilde{\Delta }_a=\omega_b$. Other parameters are provided in the text.}
\label{fig3}
\end{figure}

Figure~\ref{fig3}(a) shows the stationary optomagnonic entanglement versus the effective detunings $\tilde{\Delta}_m$ and $\tilde{\Delta}_a$. The entanglement is maximized at the detunings  $-\tilde{\Delta }_m \simeq \tilde{\Delta }_a \simeq \omega_b$, confirming our preceding analyses on the mechanism of entanglement generation, as depicted in Fig.~\ref{fig2}.  The stationary entanglement is guaranteed by the negative eigenvalues (real parts) of the drift matrix $A$. We use experimentally feasible parameters~\cite{PRAp,MA}: $\omega_m/2\pi=5$ GHz, $\omega_b/2\pi=40$ MHz, an optical wavelength $\lambda=1064$ nm, $\kappa_m/2\pi=3$ MHz, $\kappa_a/2\pi=2$ MHz, $\gamma_b/2\pi=10$ Hz, $g_m/2\pi=10$ Hz, $g_a/2\pi=1$ kHz, $T=10$ mK, and the effective coupling strengths $G_a/2\pi=4$ MHz and $G_m/2\pi=1$ MHz, corresponding to the laser power $P_L=7.51$ mW and the microwave drive power $P_m=3.08$ mW for a $5.0 \times 0.6 \times 0.3$ $\mu$m$^3$ YIG bridge~\cite{Note}. Note that a stronger optomechanical coupling strength $G_a > G_m$ is used to keep the system stable, where the optomechanical anti-Stokes process by absorbing phonons outperforms the magnomechanical Stokes process by emitting phonons.  
It is worth noting that the entanglement mechanism differs from that of Ref.~\cite{Jie18}, where a red-detuned microwave drive is applied onto the magnon mode to cool the mechanical mode close to the ground state, which is a precondition for preparing quantum states in the system.  The drive field must be {\it sufficiently strong}, such that the weak-coupling condition $G_m \ll \omega_b$ for taking the rotating-wave approximation to obtain the cooling (beam-splitter) interaction is no longer satisfied, and the counter-rotating-wave terms start to play the role, which are responsible for the parametric down-conversion and creating magnomechanical entanglement.  The red-detuned strong microwave drive used in Ref.~\cite{Jie18} plays two roles in simultaneously cooling mechanical motion and creating magnomechanical entanglement. These two roles are played, respectively, by the red-detuned driven optical cavity and the blue-detuned driven magnon mode in this work.

The optomagnonic entanglement is robust against thermal noises. We plot the entanglement versus the bath temperature in Fig.~\ref{fig3}(b), and the entanglement survives up to $T \simeq145$ mK under the parameters of Fig.~\ref{fig3}(a). For the ``mirror-on-the-surface" configuration (Fig.~\ref{fig1}), the attached mirror will cause an additional mechanical damping of the vibration mode. The additional damping, however, has a negligible influence on the entanglement. This can be seen from Fig.~\ref{fig3}(c) that the entanglement almost remains constant as the damping rate increases to $\gamma_b/2\pi \sim 10^4$ Hz, and the entanglement is still present up to a significantly large damping rate $\gamma_b/2\pi \simeq 2.5 \times 10^6$ Hz.

	

\section{Optical readout of magnonic states}\label{read}

Optical readout and transmission of localized magnonic states in solids play a crucial role in building a future magnonic quantum network~\cite{JiePRX}, and implementing remote quantum operations. Although it was proved that an arbitrary magnonic (either classical or quantum) state can be read out by using fast optical pulses~\cite{JiePRX} in cavity optomagnonic systems~\cite{Usami16,Tang16,Haigh}, the protocol suffers a low optomagnonic state-conversion efficiency because of the currently weak optomagnonic coupling and the much larger cavity decay rate. 

\begin{figure}[t]
\includegraphics[width=\linewidth]{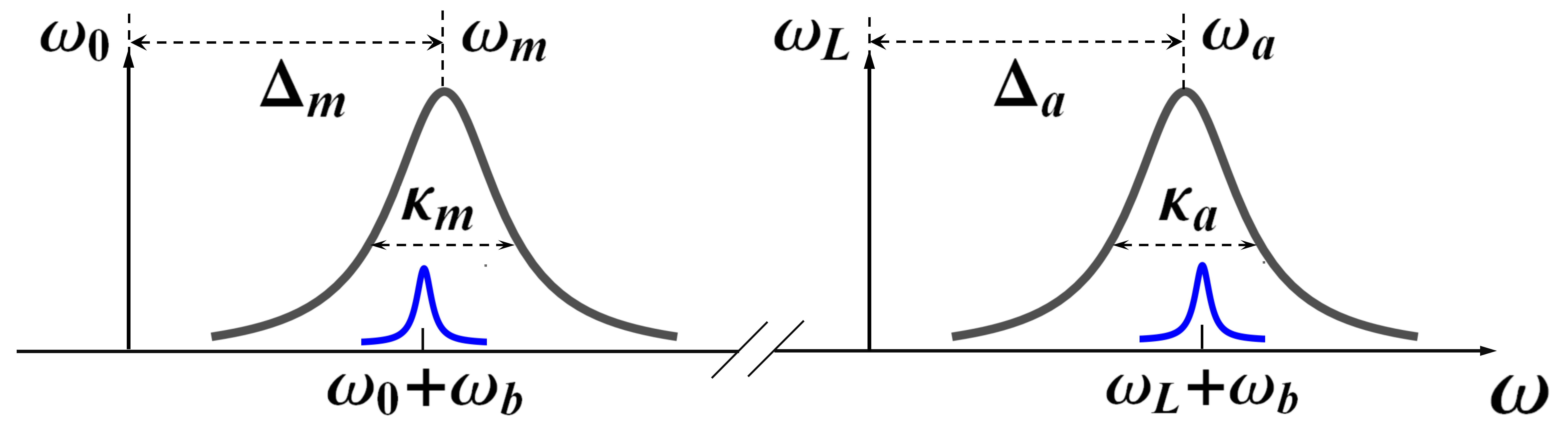}
\caption{Mode frequencies and linewidths used in the state-readout protocol. When the magnon (cavity) mode is resonant with the anti-Stokes sideband of the microwave drive field (the optical pulse) at the frequency of $\omega_0$ ($\omega_L$), the magnonic state can be transferred to phonons, then from phonons to cavity photons, and eventually to the output field of the pulse. }
\label{fig4}
\end{figure}

Here we provide a new approach using the indirectly coupled opto-magnomechanical system, which can evade the low conversion efficiency problem existing in the directly coupled optomagnonic systems. The protocol consists of two steps. We first transfer the magnonic state to the mechanical mode by activating the magnomechanical state-swap interaction realized by driving the magnon mode with a red-detuned microwave field~\cite{Jie19a,QST}, see Fig.~\ref{fig4}. When the magnomechanical system reaches a steady state, we then switch off the drive. After a short period, $ \kappa_m^{-1} < \tau_0 \ll \gamma_b^{-1}$, when the magnon excitations completely dissipate, while the mechanical state remains almost unchanged due to a much longer mechanical lifetime, we then send a weak red-detuned optical pulse to the cavity to activate the optomechanical state-swap interaction. The two successive state-swap operations (magnon-to-phonon and phonon-to-photon) transfer the magnonic state eventually to the cavity output field of the pulse, which can be readily measured by optical means, e.g., homodyne detection or state tomography. Note that in each step, we consider a two-mode model, in view of the fact that the magnomechanical coupling is weak ($g_m \ll g_a$) when the magnon drive is switched off and the magnons die out. This allows us to assume that, in the second step, the magnon mode is essentially decoupled from the optomechanical system when the pulse is applied.

The Hamiltonian of the magnomechanical subsystem is given by
\begin{align}
	\begin{split}
		H_{mb}/\hbar = \!\! \sum_{j=m,b} \omega_j j^\dagger j + \! \frac{g_m}{\sqrt{2}}  m^\dagger m \left( b+b^\dagger \right) +i\Omega \left(m^\dagger e^{-i\omega_0 t}- {\rm H.c.}\right),
	\end{split}
\end{align}
where, to express the physics more transparently, we use the operator $b$ to denote the mechanical mode, which is related to the displacement $q$ via $q=(b+b^\dagger)/\sqrt{2}$. In the frame rotating at the drive frequency $\omega_0$, we obtain the following QLEs:
\begin{align}\label{eq24}
	\begin{split}
		\dot{b}=&-i\omega_b b -\frac{\gamma_b}{2}b - i\frac{ g_m}{\sqrt{2}} m^\dagger m + \sqrt{\gamma_b}b_{in}, \\
		\dot{m}=&-i\Delta_m m -\frac{\kappa_m}{2}m -i\frac{ g_m}{\sqrt{2}} m \left( b+b^\dagger \right)+\Omega+\sqrt{\kappa_m} m_{in},
	\end{split}
\end{align}
where $b_{in}$ denotes the mechanical input noise, which obeys the correlation functions: $\langle b_{in}(t) \, b^\dagger_{in}(t') \rangle=\left[N_b(\omega_b)+1 \right] \delta(t-t')$, and $\langle b^\dagger_{in}(t) \, b_{in}(t') \rangle=N_b(\omega_b)\delta(t-t')$. Using the same linearization treatment, we get the QLEs for the fluctuations
\begin{align}
	\begin{split}
		\delta\dot{b}=&-i\omega_b\delta b-\frac{\gamma_b}{2}\delta b + \frac{G_m}{2} \left(\delta m^\dagger - \delta m \right)+\sqrt{\gamma_b}b_{in}, \\
		\delta\dot{m}=&-i\tilde{\Delta}_m\delta m-\frac{\kappa_m}{2}\delta m +\frac{G_m}{2} \left(\delta b+\delta b^\dagger \right)+\sqrt{\kappa_m}m_{in},
	\end{split}
\end{align}
where the definition of the effective coupling strength $G_m$ is the same as in Eq.~\eqref{QLEs2}. 

Moving to another interaction picture by introducing the slowly moving operators $\delta \tilde{b}=\delta b e^{i\omega_bt}$, $\delta \tilde{m}=\delta m e^{i\tilde{\Delta }_m t}$, and the noise operators $\tilde{b}_{in} =b_{in} e^{i\omega_b t}$, $\tilde{m}_{in} =m_{in} e^{i\tilde{\Delta}_m t}$, and by further taking the optimal detuning $\tilde{\Delta }_m= \omega_b$ for realizing state transfer and neglecting nonresonant fast-oscillating terms (valid when $\omega_b \gg G_m, \kappa_m, \gamma_b$), we obtain
\begin{align}\label{mbEqs}
\begin{split}
	\delta \dot{\tilde{b}} \approx &-\frac{\gamma_b}{2}\delta \tilde{b}-\frac{G_m}{2}\delta \tilde{m}+\sqrt{\gamma_b}\tilde{b}_{in}, \\
	\delta \dot{\tilde{m}} \approx &-\frac{\kappa _m}{2}\delta \tilde{m}+\frac{G_m}{2}\delta \tilde{b}+\sqrt{\kappa_m}\tilde{m}_{in},
\end{split}
\end{align}
which corresponds to an effective beam-splitter interaction Hamiltonian, accounting for the magnon-phonon state-swap operation.

We now consider a specific example where the magnon mode is prepared in a squeezed vacuum state, and see how the squeezing can be transferred to the mechanics by applying a red-detuned drive field. The magnonic squeezed vacuum can be characterized by the following squeezed noise correlations~\cite{Jie19a,QST}: $\langle \tilde{m}_{in}(t)\tilde{m}_{in}^\dagger(t')\rangle=({\cal N}+1)\delta(t-t'),\ \langle \tilde{m}_{in}^\dagger(t)\tilde{m}_{in}(t')\rangle= {\cal N} \delta(t-t'),\ \langle \tilde{m}_{in}(t)\tilde{m}_{in}(t')\rangle={\cal M}\, \delta(t-t')$, and $\langle \tilde{m}_{in}^\dagger(t)\tilde{m}_{in}^\dagger(t')\rangle={\cal M}^\ast\delta(t-t')$, where ${\cal N}=\textup{sinh}^2r$, and ${\cal M}= \textup{sinh}r\ \textup{cosh}r$, with $r$ being the squeezing parameter, characterizing the degree of the magnon squeezing.

\begin{figure}[t]
\includegraphics[width=\linewidth]{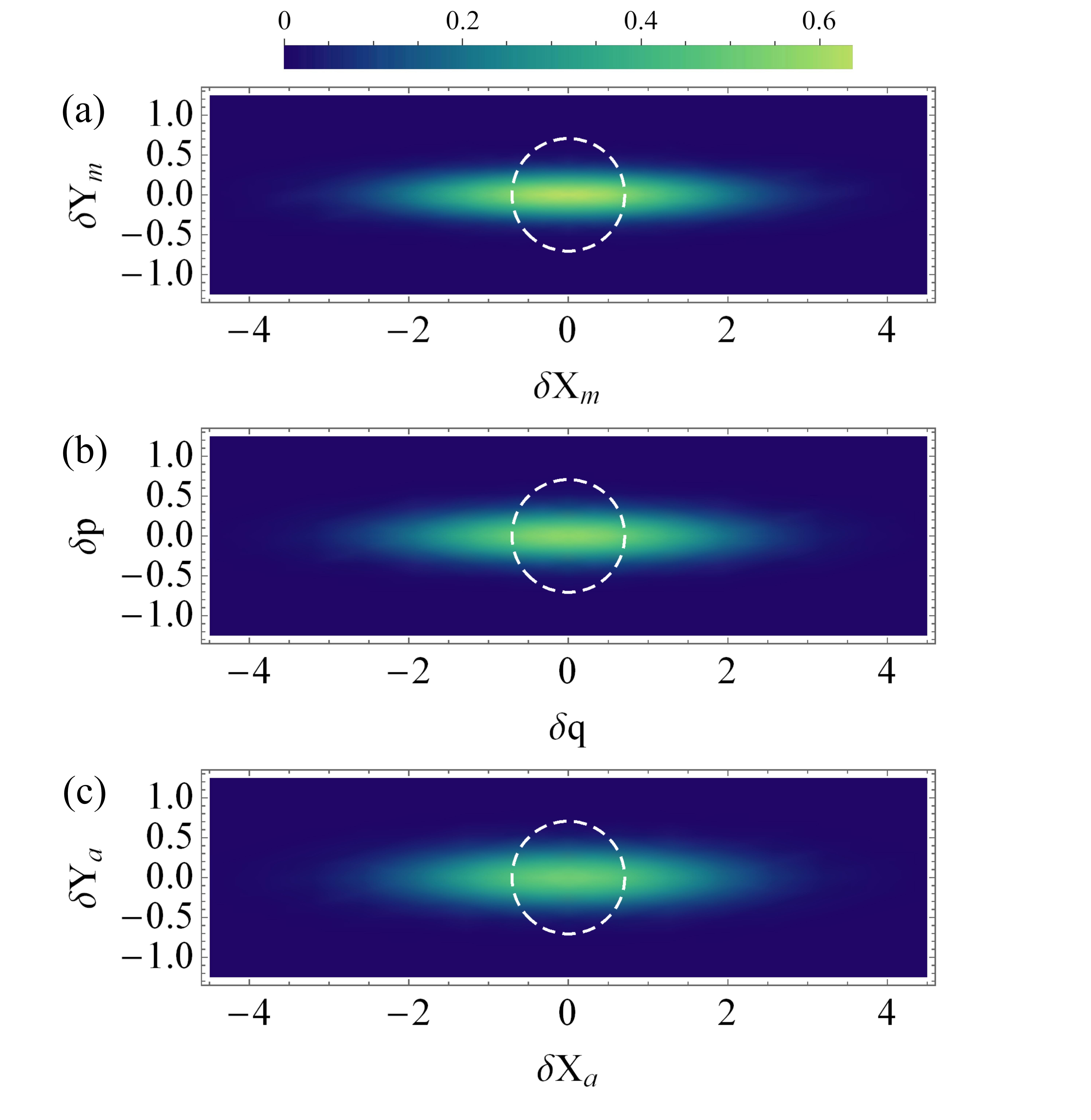} 
\caption{Wigner distribution of (a) the squeezed ($r=1$) magnon mode before applying the microwave drive, (b) the steady-state mechanical mode under the drive, and (c) the output field of the pulse after the magnon drive is turned off. The dashed white circle denotes vacuum fluctuations, below which the corresponding quadrature is squeezed. See the text for the parameters.}
\label{fig5}
\end{figure}

The CM of the mechanical mode $V_b$ can be achieved by solving the QLEs~\eqref{mbEqs}. The steady-state $V_b$ can be obtained more conveniently by solving the Lyapunov equation (see the Appendix). Given the CM $V_b$, one can calculate the Wigner function of the mechanical mode~\cite{BR}
\begin{equation}
W(u_b)=\frac{{\rm exp}(-u_b \, V_b^{-1} \, u_b^{\rm T})}{\pi \sqrt{{\rm det} \, V_b}},
\end{equation}
where $u_b=(\delta q, \delta p)$ denotes the phase-space variables associated with the fluctuations of the mechanical position and momentum. Similarly, one can get the Wigner function $W(\delta X_m, \delta Y_m)$ of the squeezed magnon mode at the initial time. The squeezed Wigner distribution $W(\delta q, \delta p)$ in phase space clearly shows the squeezing is transferred from magnons to phonons by applying a red-detuned magnon drive, c.f. Figs.~\ref{fig5}(a) and~\ref{fig5}(b). The squeezing is reduced, to some extent, due to the presence of thermal noises and dissipations of the system. We use the same magnomechanical parameters as in Fig.~\ref{fig3}, but take a smaller coupling rate $G_m/2\pi=-0.1\ \textup{MHz}$, corresponding to the microwave drive power $P_m \simeq 0.03$ mW. The fidelity in this state transfer can be calculated via the Wigner function as~\cite{Glau,Wel}
\begin{equation}
{\cal F}=\pi \int W_b(\alpha) W_m(\alpha) d^2 \alpha,
\end{equation}
where $W_j(\alpha)$ ($j=b,m$) are the Wigner function of the mechanical and magnon modes, respectively. We obtain a fidelity of ${\cal F}=0.95$ in this magnon-to-phonon state transfer.

When the system reaches its steady state, we switch off the magnon drive. After a short period, $ \kappa_m^{-1} < \tau_0 \ll \gamma_b^{-1}$, during which the magnon excitations completely decay whereas the mechanical state remains virtually unchanged, we then send a weak red-detuned pulse to the optical cavity to activate the optomechanical beam-splitter interaction, which transfers the mechanical state to the cavity output field. The pulse duration is much shorter than the mechanical lifetime, such that the mechanical damping within the pulse duration is negligible and thus can be neglected for simplicity~\cite{JiePRX}.

The Hamiltonian of the optomechanical subsystem reads
\begin{align}
H_{ab}/\hbar = \!\! \sum_{j=a,b} \omega_j j^\dagger j - \! \frac{g_a}{\sqrt{2}}  a^\dagger a \left( b+b^\dagger \right) + iE(a^\dagger e^{-i\omega_L t}-{\rm H.c.}).
\end{align}
Following the same approach used from Eq.~\eqref{eq24} to Eq.~\eqref{mbEqs}, and neglecting the mechanical dissipation, we obtain the QLEs (the tilde signs are removed for convenience)
\begin{align}
\begin{split}
	  \delta \dot{{a}} \approx &-\frac{\kappa_a}{2}\delta {a}+\frac{G_a}{2}\delta {b}+\sqrt{\kappa_a}{a}_{in},\\
	  \delta \dot{{b}} \approx & -\frac{G_a}{2} \delta {a},
\end{split}
\end{align}
which are derived under the conditions of the optimal detuning $\tilde{\Delta }_a= \omega_b$, and $\omega_b \gg G_a, \kappa_a$ for taking the rotating-wave approximation. The definition of the coupling $G_a$ is consistent with that in Eq.~\eqref{QLEs2}.

To simplify the model, we consider a flattop pulse and thus a constant coupling $G_a$ during the pulse. The pulse strength is relatively weak to have a weak coupling $G_a \ll \kappa_a$, which allows one to adiabatically eliminate the cavity, and obtain $\delta a \simeq \frac{G_a}{\kappa_a} \delta {b}+ \frac{2}{\sqrt{\kappa_a}} a_{in}$. By using the cavity input-output relation, ${a}_{out}=\sqrt{\kappa_a}\delta{a}-a_{in}$, we have
\begin{align}\label{aout-b}
\begin{split}
 a_{out} &= \sqrt{2 \GG_a} \delta b + a_{in},  \\
 \delta \dot{b} &=-\GG_a \delta{b} -\sqrt{2\GG_a}{a}_{in}, 
\end{split}
\end{align}
where $\GG_a=G_a^2/2\kappa_a$. Following~\cite{Hofer}, we define the normalized input and output temporal modes for the cavity driven by the pulse as
\begin{align}
\begin{split}
	  A_{in}(t)=&\sqrt{\frac{2\GG_a}{e^{2\GG_a t}-1}}\int_{0}^{t}e^{\GG_a \tau}{a}_{in}(\tau)d \tau , \\
	  A_{out}(t)=&\sqrt{\frac{2\GG_a}{1-e^{-2\GG_a t}}}\int_{0}^{t}e^{-\GG_a \tau}{a}_{out}(\tau)d \tau .
\end{split}
\end{align}
By integrating Eqs.~\eqref{aout-b}, we obtain the following solutions~\cite{JiePRX} 
     \begin{align}\label{Aout=b}
     	\begin{split}
     		A_{out}(t)=&\sqrt{1-e^{-2\GG_a t}}  \delta {b}(0)+e^{-\GG_a t}A_{in}(t), \\
		\delta {b}(t)=&\ e^{-\GG_a t} \delta {b}(0)-\sqrt{1-e^{-2\GG_a t}}A_{in}(t).
     	\end{split}
     \end{align}
 From the first equation of Eq.~\eqref{Aout=b}, we can get the CM of the output field of the pulse, i.e., 
 \begin{align}
 	V_{out}(t)=S V_{b}(0) + (1-S)V_{in}(t),
 \end{align}
where $S \equiv 1-e^{-2\GG_a t}$ ($0\,{<}\,S\,{<}\,1$) represents the state conversion efficiency, depending on the pulse strength and duration, $V_{in}(t)$ is the CM of the input vacuum noise, and $V_{b}(0)$ is approximately the steady-state CM $V_{b}$ of the mechanical mode obtained in the first step, i.e., $V_{b}(0) \approx V_{b}$.    For  $\GG_a t \gg 1$, $S \to 1$, and therefore $V_{out}(t) \approx V_b(0)$, which means the mechanical state is almost perfectly transferred to the output field of the pulse. 
 
Using a relatively weak coupling $G_a/2\pi=0.3\ \textup{MHz} \ll \kappa_a/2\pi=2\ \textup{MHz}$ (corresponding to the pulse power $P_L\simeq 40$ $\mu$W), and pulse duration $t=10$ $\mu$s, we obtain the state conversion efficiency $S \simeq 0.94$. The rest of the parameters are the same as in Fig.~\ref{fig3}. The Wigner function of the pulse output field is displayed in Fig.~\ref{fig5}(c). The nonunity conversion efficiency adds additional vacuum noise into the mechanical squeezed state, thus reducing the degree of squeezing, as shown by comparing Figs.~\ref{fig5}(b) and \ref{fig5}(c). 
	 
Figure~\ref{fig5} shows explicitly that the magnon squeezed state is read out in the optical pulse field via the mediation of phonons. This is realized by exploiting the magno- and optomechanical state-swap interactions in the two-step protocol. The sum of the noises added in the two processes determines the final fidelity between the magnonic state and the optical readout state, which is ${\cal F}=0.89$. Since in the second step, the mechanical dissipation is negligible during the short pulse interaction, reducing the thermal noise in the first step would play an essential role in improving the fidelity in the magnon-to-photon state transfer. This can be realized, e.g., by improving the magnomechanical coupling $G_m$ for mechanical cooling and placing the sample at a low temperature.

\section{Conclusion}\label{conc}

We propose a novel opto-magnomechanical configuration, where magnons and optical photons couple to vibration phonons by dispersive magnetostrictive and radiation-pressure interactions. We strictly derive the magnomechanical interaction Hamiltonian, and specify the condition under which the dispersive magnon-phonon coupling is dominant. We show that this opto-magnomechanical system can be adopted to prepare a {\it stationary} entangled state between magnons and optical photons, which cannot be realized in current directly-coupled optomagnonic systems. We further show that the system can also realize the optical readout of magnonic states in solids by utilizing fast optical pulses and the mediation of phonons. We expect that such a newly developed opto-magnomechanical system could find various promising applications in quantum magnonics, quantum information processing, and quantum networks.

\section*{ACKNOWLEDGMENTS}

This work has been supported by Zhejiang Province Program for Science and Technology (Grant No. 2020C01019) and the National Natural Science Foundation of China (Grant No. 11874249).

\section*{APPENDIX}

The QLEs \eqref{mbEqs} can be rewritten in the quadrature and matrix form of
	\begin{align}
		\dot{u'}(t)=R u'(t)+n'(t),
	\end{align}
where $u'(t)=\left [\delta q(t),\delta p(t),\delta X_m(t),\delta Y_m(t) \right ]^T$, and $n'(t)=[\!\sqrt{\gamma_b}\delta X_{b,in}(t),  \sqrt{\gamma_b}\delta Y_{b,in}(t),\sqrt{\kappa_m}\delta X_{m,in}(t), \sqrt{\kappa_m}\delta Y_{m,in}(t) ]^T$. For simplicity, we remove the tilde signs on the operators. The drift matrix $R$ is given by
	\begin{align}
	R=\begin{pmatrix}
		-\frac{\gamma_b}{2} & 0 & -\frac{G_m}{2} & 0 \\
		0 & -\frac{\gamma_b}{2} & 0 & -\frac{G_m}{2} \\
		\frac{G_m}{2} & 0 & -\frac{\kappa _m}{2} & 0 \\
		0 & \frac{G_m}{2} & 0 & -\frac{\kappa _m}{2} \\
	\end{pmatrix}.
	\end{align}
The diffusion matrix $Z$, which is defined as $Z_{ij}=\langle n'_i(t) n'_j(t')+n'_j(t') n'_i(t) \rangle/2$, takes the form of	
	\begin{equation}
	Z=\begin{pmatrix}
		\gamma_b(N_b+\frac{1}{2}) & 0 & 0 & 0 \\
		0 & \gamma_b(N_b+\frac{1}{2}) & 0 & 0 \\ 
		0 & 0 & Z_{m}^{11} \,& Z_{m}^{12} \\ \,\,\,
		0 & 0 & Z_{m}^{21} \,& Z_{m}^{22} \\  
	\end{pmatrix},
	\end{equation}
where $Z_{m}^{11}=\frac{\kappa_m}{2}(2N+1+M+M^\ast)$, $Z_{m}^{12}=Z_{m}^{21}=\frac{i \kappa_m}{2}(M^\ast-M)$, and $Z_{m}^{22}=\frac{\kappa_m}{2}(2N+1-M-M^\ast)$. The steady-state $4\times4$ CM $V_{mb}$ of the magnomechanical system can then be achieved by solving the Lyapunov equation
\begin{equation}
	R V_{mb}+V_{mb} R^T= -Z.
\end{equation}
Once the $V_{mb}$ is obtained, one then gets the $2\times2$ CM of the mechanical (magnon) mode by removing the irrelevant 
rows and columns in $V_{mb}$.


\begin{thebibliography}{99}
\bibitem{Note}
The time average of energy density is $\rho_E=H_d^2/2\mu_0$ ($\mu_0$ is the vacuum magnetic permeability). Thus, the microwave drive power $P_m=\rho_E Ac$, where $c$ is the speed of an electromagnetic wave propagating in vacuum, and $A$ is the cross-sectional area of the YIG micro bridge ($A=L\times W$, with $L$ and $W$ being the length and width of the micro bridge). The drive magnetic field is applied perpendicular to the cross-section (Fig.~\ref{fig1}). Therefore, we obtain the relation between the drive magnetic field $H_d$ and the power $P_m$ via $H_d=\sqrt{2\mu_0 P_m/(LWc)}$.
\end{thebibliography}
\end{document}